\documentclass[apjl]{emulateapj}
\usepackage{multirow}

\def\Msun{M$_{\odot}$}

\def\Mstar{$M_{\star}$}

\def\HII{H\,{\scriptsize II}}
\def\ha{${\rm H}\alpha$}
\def\hb{${\rm H}\beta$}
\def\oiii{[O\,{\scriptsize III}]}
\def\oii{[O\,{\scriptsize II}]}
\def\nii{[N\,{\scriptsize II}]}
\def\lya{${\rm Ly}\alpha$}
\def\hayashi12{H12}
\slugcomment{Accepted for publication in the ApJ Letters on 13 July 2016.}
\shorttitle{Main sequence in the USS1558-003 protocluster}
\shortauthors{Hayashi M. et al.}

\begin{document}

\title{Enhanced Star Formation of Less Massive Galaxies in a Proto-Cluster at z=2.5}

\author{Masao Hayashi\altaffilmark{1}, 
  Tadayuki Kodama\altaffilmark{1,3}, 
  Ichi Tanaka\altaffilmark{2},
  Rhythm Shimakawa\altaffilmark{3},
  Yusei Koyama\altaffilmark{2},  
  Ken-ichi Tadaki\altaffilmark{4}, 
  Tomoko L. Suzuki\altaffilmark{3}, 
  and 
  Moegi Yamamoto\altaffilmark{3}
}

\affil{$^{1}$Optical and Infrared Astronomy Division, National
  Astronomical Observatory, Mitaka, Tokyo 181-8588, Japan; masao.hayashi@nao.ac.jp}
\affil{$^{2}$Subaru Telescope, National Astronomical Observatory of
  Japan, 650 North A'ohoku Place, Hilo, HI 96720, USA} 
\affil{$^{3}$Department of Astronomical Science, SOKENDAI (The
  Graduate University for Advanced Studies), Mitaka, Tokyo 181-8588} 
\affil{$^{4}$Max-Planck-Institut f$\ddot{\rm u}$r Extraterrestrische
  Physik, Giessenbachstrasse, D-85748 Garching, Germany} 

\begin{abstract}
We investigate a correlation between star-formation rate (SFR) and
stellar mass for \ha\ emission line galaxies (HAEs) in one of the
richest proto-clusters ever known at $z\sim2.5$, USS~1558-003
proto-cluster. This study is based on a 9.7-hour narrow-band imaging
data with MOIRCS on the Subaru telescope. We are able to construct a
sample, in combination with additional $H$-band data taken with WFC3
on Hubble Space Telescope (HST), of 100 HAEs reaching the
dust-corrected SFRs down to 3 \Msun\ yr$^{-1}$ and the stellar masses
down to $10^{8.0}$ \Msun. We find that while the star-forming galaxies
with $\ga10^{9.3}$ \Msun\ are located on the universal SFR-mass main
sequence irrespective of the environment, less massive star-forming
galaxies with $\la10^{9.3}$ \Msun\ show a significant upward scatter
from the main sequence in this proto-cluster. This suggests that some
less massive galaxies are in a starburst phase, although we do not
know yet if this is due to environmental effects. 
\end{abstract}



\section{Introduction}
Since the last decade, the question of a positive correlation between
SFR and stellar mass in star-forming galaxies (SFGs), which is called
the main sequence (MS) of SFGs, has been one among hot topics in the
field of galaxy evolution \citep[e.g.,][]{Noeske2007,Daddi2007,Elbaz2007}.
The tight correlation provides us with perspectives of how the SFGs
evolve over cosmic time: They spend most of their lifetimes on the
sequence and evolve along the MS. However, a small fraction of them
shows starburst activities, and they deviate upwards from the MS
\citep{Rodighiero2011}. 

During the course of hierarchical structure formation, galaxy
evolution is expected to proceed in different ways in different
environments. It is suggested that such environmental effects are more 
preferentially seen in satellite galaxies rather than in central
galaxies at $z<1$ \citep[e.g.,][]{Peng2012,Kovac2014}.
Some environment-dependent processes such as galaxy
interactions/merging and gas inflows/outflows can alter the
star-formation activity in galaxies, either boosting it or truncating
it. Understanding the physical mechanisms of these processes is of
vital importance to reveal the origin of early-type galaxies and the
strong environmental dependence of galaxy properties seen in the
present-day Universe.  

With this motivation, we have been conducting a systematic project
called MAHALO-Subaru (MApping H-Alpha and Lines of Oxygen with Subaru;
\citet{Kodama2013}) and mapping star-formation activities over a wide
range of environments and across cosmic times, in particular at
$1.5 \lesssim z \lesssim 2.5$, where clusters of galaxies are just
assembling and galaxies are forming vigorously therein. The project
has shown that integrated SFR per dynamical mass in cluster core
increases dramatically with redshift up to $z\sim2.5$
\citep{Shimakawa2014}. However, \citet{Koyama2013b} show that the
location of the MS of SFGs in a proto-cluster (PKS1138--262) at
$z\sim2$ is not different from that in the general field at similar
redshifts, although the distribution of galaxies along the MS is
skewed to higher SFRs and stellar masses in high density regions 
probably due to biased, more advanced galaxy formation there. Since
our analyses have been limited to relatively massive galaxies
($\ga10^{9.5}$ \Msun) so far, we want to extend such study to an even
less massive regime.

For this purpose, we target a proto-cluster around the USS~1558-003
radio galaxy at $z=2.53$. \citet[][hereafter \hayashi12]{Hayashi2012}
have already reported a narrow-band \ha\ emission line survey as a
part of MAHALO-Subaru. This previous observation has identified as
many as 68 HAEs associated with the proto-cluster within a 27
arcmin$^2$ field of view (FoV). It shows a linear filamentary
structure which hosts three dense groups of HAEs. The richness and
high density make it a unique proto-cluster target at $z>2$ for us to
investigate the early environmental effects in the galaxy-formation
phase.  

To access less massive galaxies in the cluster, we have conducted very
deep follow-up observations: One is three times deeper narrow-band
\ha\ imaging with Subaru Telescope. Another is deep HST/WFC3 imaging
at near-infrared.  In this {\it Letter}, based on these new unique
imaging data-sets, we report the first intriguing discovery of the
nature of the less massive SFGs ($\la10^{9.5}$ \Msun) in this
proto-cluster USS~1558-003 at $z=2.53$, which has become accessible
only with these deep observations. 

Magnitudes are presented in the AB system \citep{Oke1983},
cosmological parameters of ${\rm H}_0=70$ km s$^{-1}$ Mpc$^{-1}$,
$\Omega_{m}=0.3$ and $\Omega_{\Lambda}=0.7$, and
\citet{Chabrier2003} initial mass function, are
adopted throughout the letter.

\section{Data}
\label{sec:data}

\begin{deluxetable*}{ccccccc}
\tabletypesize{\scriptsize}
\tablecaption{The optical and near-infrared images.}
\tablewidth{0pt}
\tablehead{
\colhead{\multirow{2}{*}{Filter}} & 
\colhead{\multirow{2}{*}{Instrument/Telescope}} &  
\colhead{Integration time} & \colhead{Limiting mag.\tablenotemark{a}} &
\colhead{PSF} & \colhead{\multirow{2}{*}{Proposal ID}} \\
\colhead{}  & \colhead{} &
\colhead{(minutes)} & \colhead{(5$\sigma$)} & 
\colhead{(arcsec)} & \colhead{} & \colhead{} 
}
\startdata
$B$    & Suprime-Cam/Subaru & 80  & 27.51 & 0.70 & S10B-028 \\
$r'$   & Suprime-Cam/Subaru & 90  & 27.24 & 0.63 & S10B-028 \\
$z'$   & Suprime-Cam/Subaru & 55  & 26.03 & 0.66 & S10B-028 \\
$J$    & MOIRCS/Subaru & 191  & 24.85 & 0.55 & S10B-028, S15A-047 \\
$H_{160}$ & WFC3/HST & 87 & 27.46 & 0.21 & GO-13291 \\
$H$    & MOIRCS/Subaru & 45  & 23.78 & 0.47 & S10B-028 \\
$K_s$  & MOIRCS/Subaru & 207  & 24.49 & 0.60 & S10B-028, S15A-047 \\
NB2315 & MOIRCS/Subaru & 583 & 23.90 & 0.52 & S10B-028, S15A-047
\enddata
\tablecomments{The FWHMs of point spread function (PSF) in all the
  Subaru images are matched to 0.67\arcsec\ finally, except for the
  $B$-band image which has a FWHM of 0.70\arcsec.}  
\tablenotetext{a}{The limiting magnitudes are measured with a
  1.2\arcsec\ diameter aperture, except for 0.4\arcsec\ aperture for
  WFC3/HST image.}   
\label{table:data}
\end{deluxetable*}

\begin{figure}
  \includegraphics[width=0.5\textwidth]{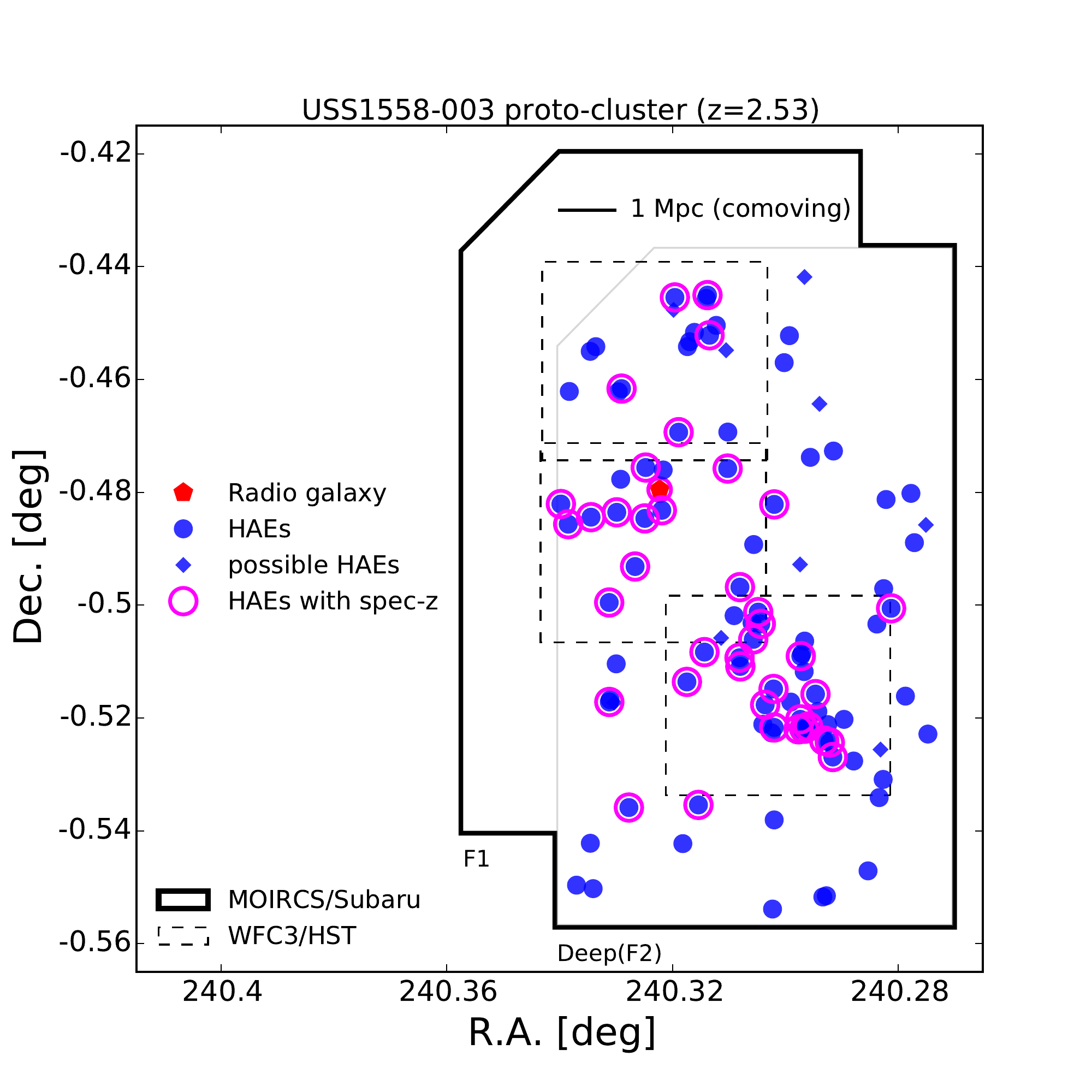}  
  \caption{
    The FoVs of our deep imaging data by MOIRCS/Subaru (solid
    line; 39.0 arcmin$^2$) and by WFC3/HST (dashed line; 4.9 arcmin$^2$
    $\times$ 3 fields). The red pentagon indicates the USS1558-003 radio
    galaxy at $z=2.53$. \ha\ emitters we selected in \S
    \ref{sec:catalog} are shown by blue circles. \ha\ emitters with
    spectroscopic redshift are marked with magenta open
    circles. \label{fig:map}}    
\end{figure}

Since many of the imaging data used in this {\it Letter} are already
published in \hayashi12, we mainly describe the new additional data
obtained after \hayashi12. Data in $J$, $K_s$, and NB2315 are updated
by further deep observations with MOIRCS/Subaru
\citep{ichikawa2006,suzuki2008} at a single pointing whose FoV
corresponds to the field called `F2' in \hayashi12 (S15A-047, PI:
T.~Kodama). During the observations from April 30 to May 6 in 2015,
the weather was fine and the sky condition was photometric. Most of
the frames were taken under the seeing condition of better than
0.6\arcsec.  Combined with the \hayashi12 data, the total integration
times sum up to 3.18, 3.45, and 9.72 hours in $J$, $K_s$, and NB2315
(Table~\ref{table:data}), respectively. All of the $J$, $K_s$, and
NB2315 data are re-reduced in a standard manner using the data
reduction package for MOIRCS (MCSRED ver.20150619\footnote{\url{http://www.naoj.org/staff/ichi/MCSRED/mcsred.html}}
by I.~Tanaka). The details of the reduction are shown in our
forthcoming main paper. 

Observations with WFC3/HST in $F160W$ are conducted on July 5 and 9 
in 2014 (GO-13291, PI: M.~Hayashi). Three pointings with WFC3 are
required to cover the structures found in \hayashi12
(Figure~\ref{fig:map}). Since it took two orbits for the observations
in each pointing, the integration times are 5224 sec in total 
(Table~\ref{table:data}). The reduction is carried out in a standard
manner with pipeline. Using the task, {\sc multidrizzle}, the pixel
scale is changed to 0.06 arcsec per pixel \citep{Koekemoer2011}. 

To summarize, the data-set that we use consists of six
broad-band data ($B,r',z',J,H$ and $K_s$) and the narrow-band (NB2315)
data with Subaru, and the $H_{160}$ data with HST/WFC3.
The $J$, $K_s$, and NB2315 images are deeper by 0.67, 0.84 and 0.89
mag.~than those in \hayashi12 (Table~\ref{table:data}). 

\section{Catalog}
\label{sec:catalog}

\begin{figure*}
\begin{tabular}{ccc}
\hspace{0.2cm}
\includegraphics[width=0.33\textwidth]{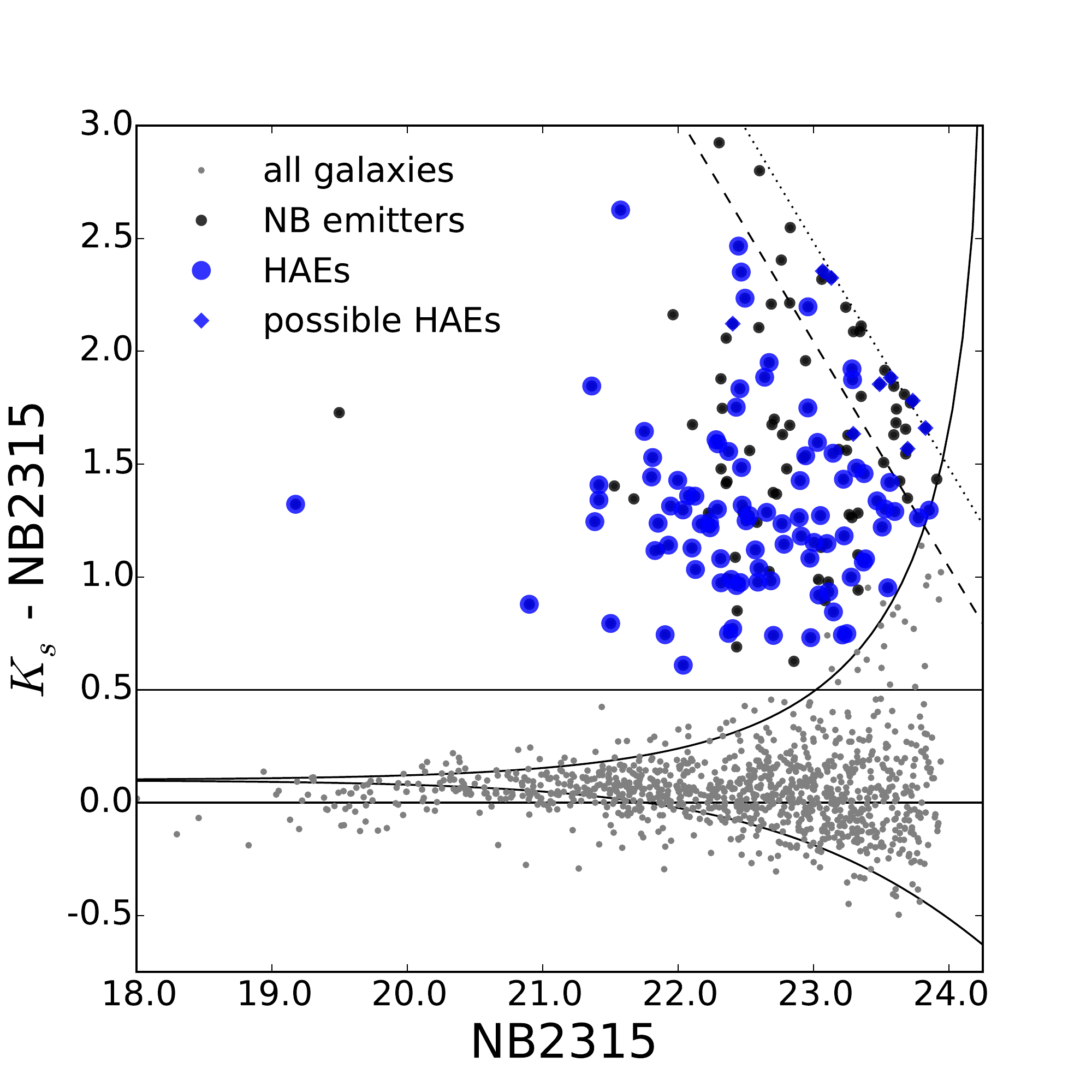} &
\hspace{-0.5cm}
\includegraphics[width=0.33\textwidth]{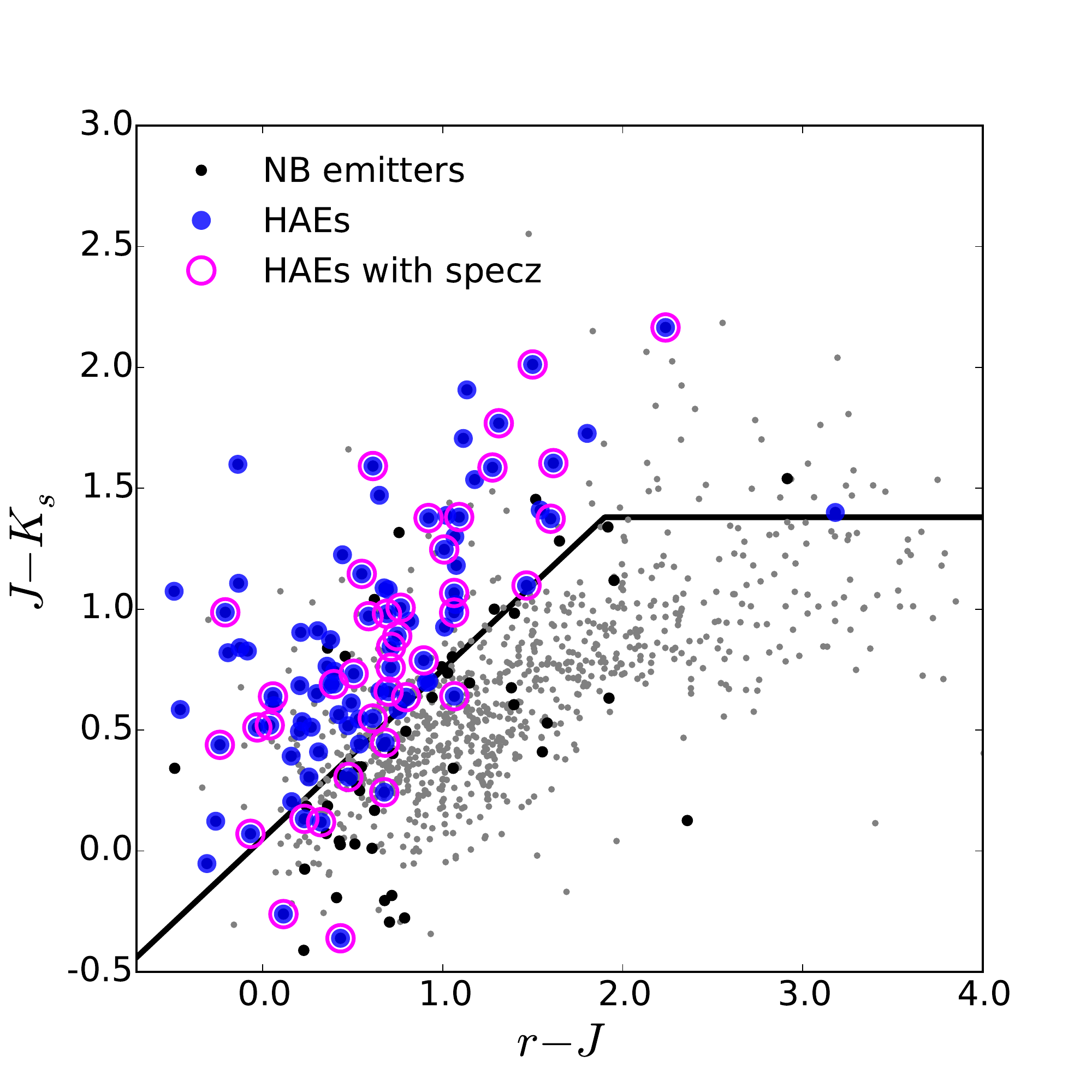} &
\hspace{-0.5cm}
\includegraphics[width=0.33\textwidth]{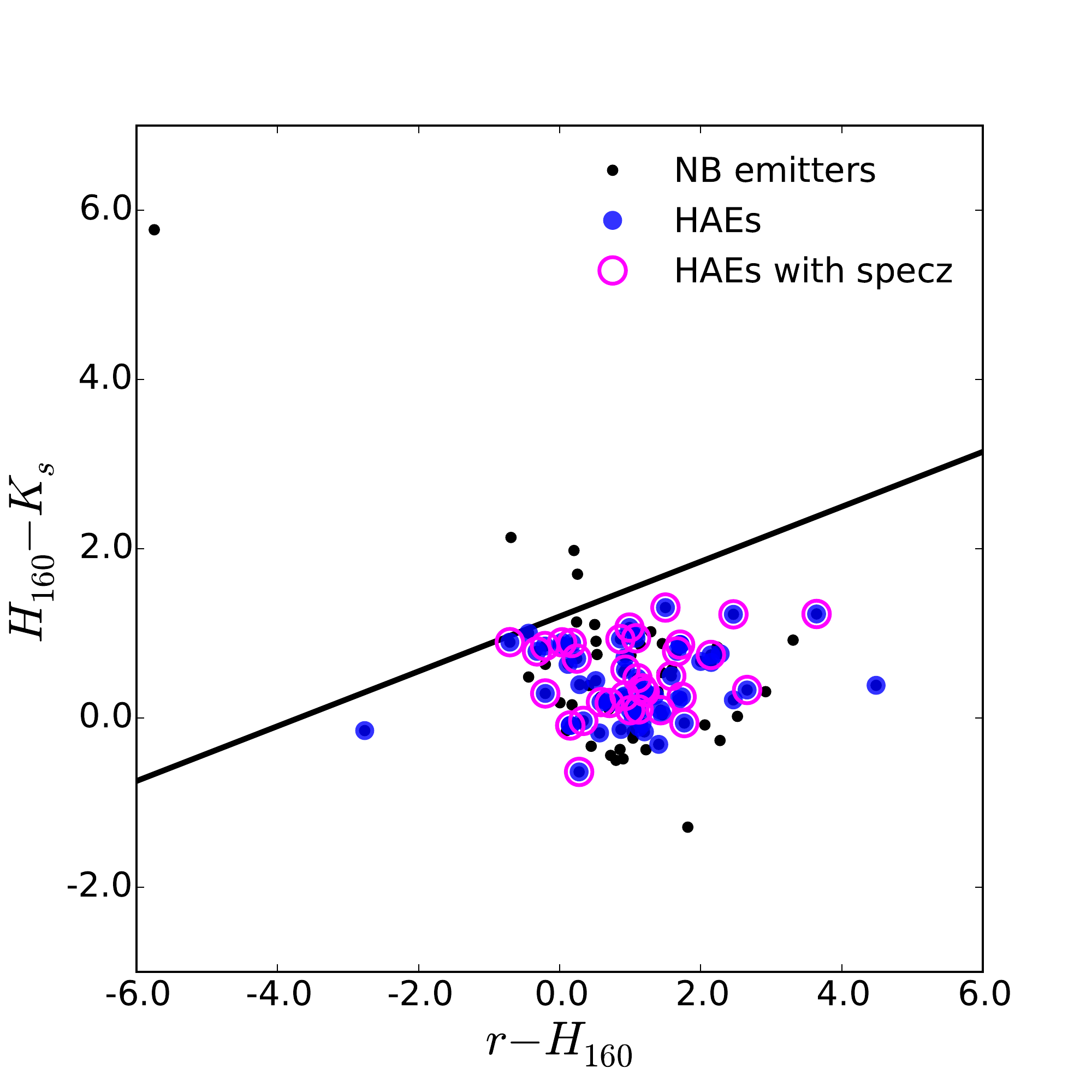}
\end{tabular}
\caption{
  (left) The color--magnitude diagram of $K_s$$-$NB2315 versus
  NB2315. The two solid curves show the boundaries of 3$\sigma$ excess colors.
  The dashed (dotted) lines show the 3$\sigma$ (2$\sigma$) limits in
  $K_s$$-$NB2315 color. The blue-filled circles represent the probable
  member HAEs at $z\approx2.53$, while blue-filled diamonds represent
  the possible HAEs (see text). The black dots located above the solid
  curve indicate the NB2315 emitters, which are not selected as HAEs
  by our color criteria.   
  (middle) The color--color diagram of $r'-J$ versus $J-K_s$. The black
  solid line shows the boundaries of our selection criteria to
  identify HAEs among the NB2315 emitters (\hayashi12). The symbols are
  the same as in the left panel, but the  
  HAEs with spectroscopic redshift are marked with magenta open
  circles. Most of the HAEs with spectroscopic redshifts that do not
  meet the $rJK_s$ criteria were faint in $J$ and $K_s$, and thus
  detected only at $<$5$\sigma$ on the images in \hayashi12.
  Now that the much deeper $J$ and $K_s$ images are available,
  the colors are determined more reliably and they turn out to be
  out of our selection criteria.
  (right) The color--color diagram of $r'-H_{160}$ versus
  $H_{160}-K_s$ to discriminate \hb\ and \oiii\ emitters at $z\sim3.6$
  from HAEs at $z\approx2.5$. The symbols are the same as in the left and
  middle panels. 
  \label{fig:selection}}
\end{figure*}

The photometric catalog is updated by taking the same procedures
on the latest data set as in \hayashi12. However, since the HST image
has much better PSF than that of the Subaru images
(Table~\ref{table:data}), source detection and photometry on the HST
image are conducted independently. The HST and Subaru photometric
catalogs are then combined by matching the detected objects within a
0.5\arcsec\ radius circle.

A procedure to select emission line galaxies is basically the same as
in \hayashi12, but we redo the selection based on the updated catalog.
First, we extract galaxies with more than 3$\sigma$ excess in
$K_s$-NB2315 color (left panel of Figure~\ref{fig:selection}). Note
that a correction of 0.1 mag.~in $K_s$-NB2315 color is required as
a color term in estimating a continuum level underneath the
\ha\ emission line, because there is a difference of 0.163\micron\ in
the effective wavelengths between $K_s$ and NB2315 filters. We also
apply the criterion of $K_s$--NB2315$>$0.50. These criteria allow us to
select 171 NB2315 emitters with line fluxes larger than
1.1$\times$10$^{-17}$ erg s$^{-1}$ cm$^{-2}$ and equivalent widths in
the observed frame larger than 66\AA. The limiting line flux
corresponds to L(\ha+\nii)=$5.8\times10^{41}$ erg s$^{-1}$ which turns
out to be 2.4 times deeper than that in \hayashi12.  It can be
converted to a dust-free SFR of 2.2 \Msun\ yr$^{-1}$
\citep{kennicutt1998}, where the contribution of \nii\ to the line
flux measured with NB, i.e.~\nii/(\ha+\nii), is assumed to be 0.25
\citep{Sobral2013}.     

To eliminate lower-$z$ contaminant lines from the NB2315 emitter
sample, we use the same color selection criteria as in \hayashi12
on the color--color diagram, $r'-J$ versus $J-K_s$ (middle panel of
Figure~\ref{fig:selection}). However, some emitters are not securely
detected in $r'$, $J$, or $K_s$, and we notice that almost all such
galaxies are very faint in $J$ and $K_s$. For many cases, we are thus 
not able to classify them according to color selection. We regard them
as `possible \ha\ emitters'. This may be justified by our
Suprime-Cam/Subaru observation with NB428 narrow-band filter targeting
\lya\ emitters at the same redshift, which shows that most of the
\lya\ emitters are not detected or very faint ($<$3$\sigma$) in $J$
and/or $K_s$ (Shimakawa et al.~submitted). 

We have also found three contaminant \oiii\ emitters at $z\sim3.6$
in the follow-up spectroscopy \citep{Shimakawa2014,Shimakawa2015b}.
This suggests a necessity of applying an additional color selection
to distinguish \oiii\ and \hb\ at $z\sim3.6$ from \ha\ at $z\sim2.5$.
We thus utilize the $r'-H_{160}$ versus $H_{160}-K_s$ diagram as well
for galaxies whose HST $H_{160}$ data are available. 
The deep HST $H_{160}$ photometries correspond to a bluer side of the
Balmer/4000\AA\ break for galaxies at $z\sim3.6$.
We set the boundary at ($r'-H_{160}$) = 0.324 ($H_{160}-K_s$) + 1.2
based on the colors of the spectroscopically confirmed HAEs to
distinguish the two populations (right panel of Figure~\ref{fig:selection}).
The slope of the criterion is parallel to a reddening vector estimated
from the dust extinction curve of \citet{Calzetti2000}. We note that
color tracks modelled using the \citet{BC03} stellar population
synthesis code also support the $rHK_s$ selection criterion.
The $rHK_s$ color selection identifies four emitters as galaxies at
$z\sim3.6$. Furthermore, $B$-band magnitudes of \oiii\ emitters at
$z\sim3.6$ are more sensitive to attenuation by neutral hydrogen in
the intergalactic medium than HAEs at $z\sim2.5$
\citep{Madau1995}. Non-detection or faint magnitude in $B$ imply that
the galaxies are more likely to be located at $z\ga3$. Thus, among the
HAEs without spectroscopic confirmation and possible HAEs, 15 emitters
without a detection at more than $2\sigma$ in $B$ or with $B-r'>1.39$
are excluded from the samples. Note that all of the confirmed HAEs,
except for one galaxy, have $B-r'<1.39$. We also note that 98\% of the
HAEs are detected in $r'$-band at 2$\sigma$ level, which suggests that
there is little contamination of \oii\ emission at $z\sim5.2$.       

Consequently, by applying these color selections, we select 91 HAEs
and nine possible HAEs in total (Figures~\ref{fig:map} and
\ref{fig:selection}). Hereafter, we treat both HAEs and possible HAEs
as SFGs at $z\sim2.5$. Comparing with the sample of
\hayashi12, the number of HAEs increases from 68 to 100, due to
deeper NB2315 and $K_s$ images available in this study. 

\section{Main sequence of star-forming galaxies}
\label{sec:MS}

\begin{figure}
\includegraphics[width=0.5\textwidth]{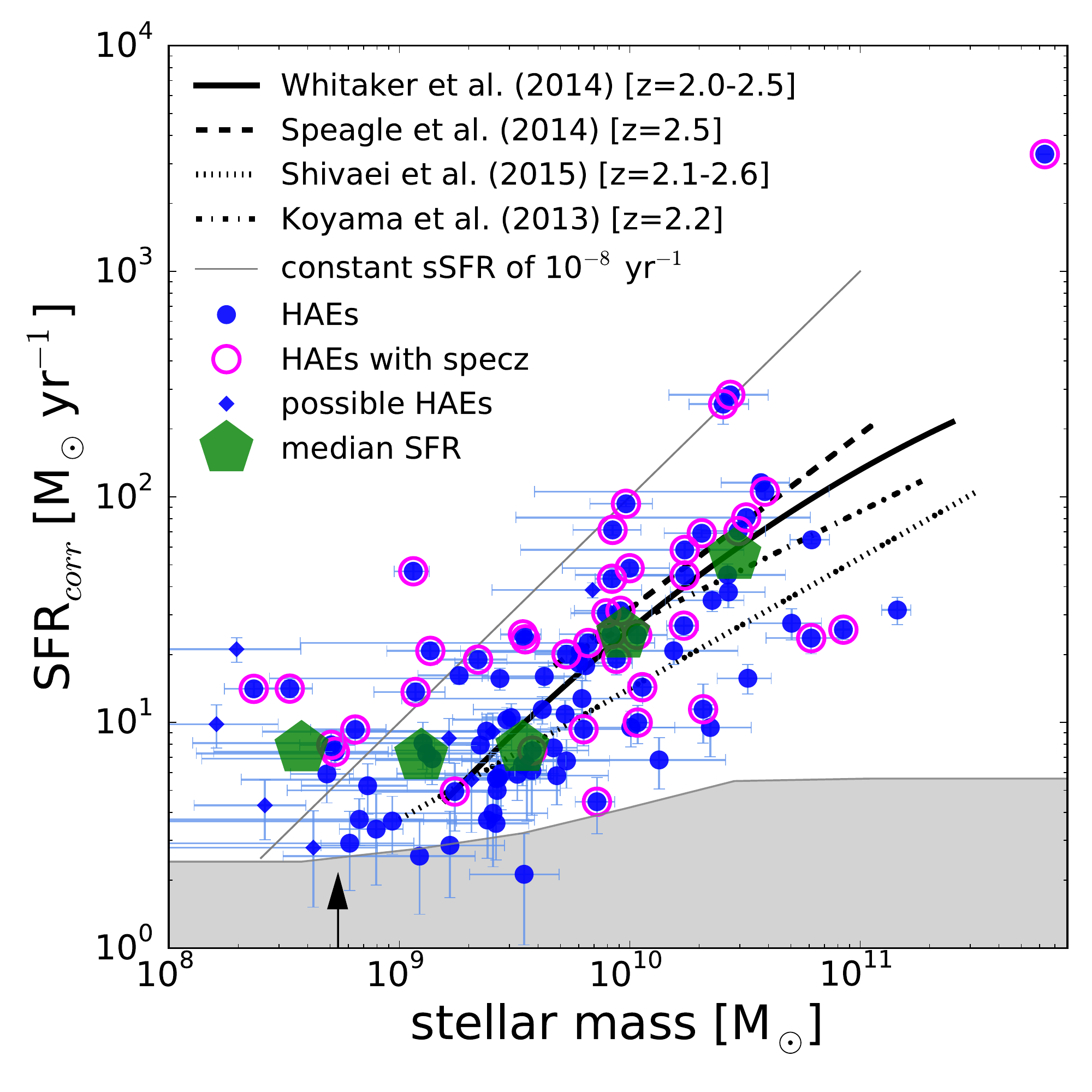}  
\caption{ Main sequence of HAEs in the USS1558-003 proto-cluster at
  $z=2.53$. The symbols are the same as in Figure~\ref{fig:selection}.
  The error bars are derived from the 1$\sigma$ photometric error.
  The uncertainties of stellar masses derived from the SED fitting
  are estimated from a standard deviation of 100 iterations.
  Pentagons show the median SFRs in each mass bin. The gray region shows
  the SFRs under the limit reachable. The curves are MSs from the
  literature \citep{Whitaker2014,Speagle2014,Shivaei2015,Koyama2013b}. 
  A gray line shows a constant specific SFR of $10^{-8}$ yr$^{-1}$,
  and an arrow roughly shows the stellar mass limit which is estimated
  with the 3$\sigma$ limiting magnitude in $K_s$ and $J-K_s$ color of 0.36
  (\hayashi12). 
  \label{fig:MS}}
\end{figure}

We estimate stellar masses of the HAEs by fitting a library of
evolutionary stellar population synthesis models
\citep[GALAXEV,][]{BC03} to spectral energy distribution (SED) with
the six broad bands. The SEDs are hardly affected by nebular emission
lines, because none of the strong lines but \oii\ enter the broad bands.    
In modelling, we fix the redshift to the spectroscopic one, if
available. Otherwise, we use the redshift of the radio galaxy 
($z=2.53$). We assume exponentially declining and constant
star-formation histories. Acceptable ages are chosen from 50 Myr to
the age of the Universe at that redshift. Stellar metallicities of
$Z_\odot$ or $0.4 Z_\odot$ are applied, which are consistent with
gas-phase metallicities of the HAEs measured by \citet{Shimakawa2015a}.  
A dust extinction curve by \citet{Calzetti2000} is assumed and
$E(B-V)$ ranges from 0.0 to 3.0.
Even if galaxies are faint in $J$ or $K_s$, thanks to the deep
$H_{160}$ and optical bands, a model SED is determined for most of the
HAEs. However, 10 HAEs are not fitted by any of the model SEDs. In
those cases, we estimate the stellar masses based on their $K_s$
magnitudes corrected for the mass-to-luminosity ratio
(\Mstar/$L_{K_s}$) measured from their $J-K_s$ colors
\citep[\hayashi12,][]{Kodama1998}. We note that for the galaxies with
SEDs available, the stellar masses derived from $K_s$ and $J-K_s$ are
consistent with those determined by the SED fitting. 

Next, we estimate SFRs of the HAEs using the \ha\ luminosities derived
from the narrow-band (NB2315) imaging (see \hayashi12 for more
details). The contribution of \nii\ is removed from the line flux by
assuming the relation between the ratio of \nii/\ha\ and the
rest-frame equivalent width of EW$_0$(\ha+\nii) given by
\citet{Sobral2013}. If the spectroscopic redshift is available, we
correct the \ha\ flux for the transmission of the NB filter at the
wavelength of the line. 
We then correct for dust extinction using the calibration by
\citet{Koyama2015} which estimates A(\ha) from the observed
SFR(UV)/SFR(\ha) ratio and the stellar mass, where the
$\sigma_{\rm rms}$ of A(\ha) in the calibration is 0.282 mag
\citep{Koyama2015}. We use $r'$-band magnitudes (i.e, rest-frame
1795\AA) to estimate the rest-frame UV luminosity densities. The
intrinsic \ha\ luminosities thus derived are converted to SFRs using
the \citet{kennicutt1998} calibration.  

Figure \ref{fig:MS} shows a positive correlation between SFR and
stellar mass for the 100 HAEs, confirming the existence of the MS of
SFGs. The shaded region in gray in Figure \ref{fig:MS} shows our limit
in SFR. This limiting SFR indicates that our data are deep enough for
us to discuss the MS for SFGs fully down to stellar masses of
$\sim10^{10}$ \Msun\ and we can still reach the upper half of the MS
galaxies down to $\sim10^{9}$ \Msun. At the lowest mass regime of
$\la10^{9}$ \Msun, we would no longer be able to access the MS
galaxies if the sequence is extrapolated to the low mass, and we could
see only the SFGs with enhanced star-formation activities above the
MS.  

\section{discussion}
\label{sec:discussion}
Until now, there have been a number of previous studies that
investigated the MS of SFGs at $z\ga2$ \citep[e.g.,][]{Dunne2009,Karim2011,Reddy2012}.
Figure \ref{fig:MS} also compares the MSs derived from the previous
studies. Our HAEs with stellar masses of $\ga10^{10}$ \Msun\ seem to
be located right on the previously measured MSs in the literature
\citep{Whitaker2014,Shivaei2015,Speagle2014}. Since these
previous studies mainly look at galaxies in the general fields, the
agreement suggests that SFGs in the proto-cluster at $z\sim2.5$ share
the same MS as field SFGs at similar redshifts, consistent with the
previous studies \citep{Koyama2013b, Cooke2014}. 

On the other hand, at lower stellar masses ($<10^{9.3}$ \Msun) in the
proto-cluster, there are several galaxies that are significantly
up-scattered above the MSs. If such small mass galaxies all follow the
same extrapolated MS, they would be all located below our detection
limit and we would not see any of them, contrary to what we actually see.
Although we cannot discuss the exact locations of the MS at
$<10^{9.3}$ \Msun\ due to incompleteness, we argue that there are at
least some HAEs that are significantly deviated upward from the MS.
Those more than 10 up-scattered HAEs at $<10^{9.3}$ \Msun\ have
exceptionally large specific SFRs (sSFR=SFR/\Mstar) above 10$^{-8}$
yr$^{-1}$ as shown by the light solid line in Figure~\ref{fig:MS}.
This indicates that their inferred ages (timescales of star formation)
are smaller than 10$^8$ years, and they are young starbursting
galaxies just being formed. Note that this result is not affected by
dust corrections, because the amount of dust correction is
progressively lower for less massive galaxies \citep{GarnBest2010,Koyama2015}. 
In fact, the inferred A(\ha) for almost all of the galaxies with
$<10^{9}$ \Msun\ are smaller than 0.2 mag.
Even if we use the rest-frame UV luminosities to derive SFRs of the
HAEs instead of \ha\ luminosities, we also find the existence of HAEs
above the MS at the faint end. 

Our results suggest that while the majority of massive galaxies are
already settled in a secular evolution phase and are thus found on the
MS, some less massive galaxies are in a starburst phase and they are
significantly up-scattered from the MS.
This may be consistent with the down-sizing scenario of mass-dependent
galaxy evolution \citep[e.g.,][]{Cowie1996,Bundy2006,Muzzin2013}, or
since they are located in a dense proto-cluster, they may be
experiencing some influences from the surrounding environment such as
galaxy-galaxy interactions. We do not know, however, if this trend is
seen only in high density regions or it is a common feature of less
massive SFGs (traced by \ha) irrespective of environment.  

The SED fitting described in \S \ref{sec:MS} indicates that the
youngest age of $<10^8$ yr is preferred for galaxies with stellar mass
less than $10^{9}$ \Msun, which is again consistent with the less massive
HAEs having sSFR of $>10^{-8}$ yr$^{-1}$.  This also supports our
interpretation that they are young, star-bursting galaxies during the
vigorous formation/assembly epoch of a rich galaxy cluster.

\citet{Cooke2014} show the lack of galaxies with stellar mass less
than $\sim10^{10}$ \Msun\ in a proto-cluster at $z=2.49$, and argue
that it is possibly due to either a large dust extinction of less
massive galaxies or the earlier formation of massive
galaxies. However, our results show that there are SFGs on the MS down
to stellar mass of $10^{9.3}$ \Msun\ and that even at lower mass bin
there are SFGs with SFRs comparable to those of more massive galaxies
with $10^{10}$ \Msun, which are not in agreement with the results by
\citet{Cooke2014}. Compared to the USS1558, the proto-cluster
discussed in \citet{Cooke2014} is not very rich, although it shows
some overdensity in contrast to the general fields. Therefore, the
discrepant result between this {\it Letter} and \citet{Cooke2014}
could be due to the intrinsic diversity of the properties of
proto-clusters at $z\sim2.5$. However, to address this issue, it is
essential for us to investigate a much larger sample of proto-clusters.     

The existence of the less massive HAEs with $<10^{9.3}$
\Msun\ up-scattered above the MS may imply that a scatter around the
MS increases at lower stellar masses. Diversity of star-formation
history in early phase of galaxy evolution and/or sensitivity to the
fluctuation of starburst activity at short time scales in individual
\HII\ regions could cause the increased scatter.  
Another possible remaining issue is a metallicity dependence of the
\ha\ luminosity \citep[e.g.,][]{Bicker2005,Dopita2006}.
A lower stellar metallicity would result in a higher stellar
temperature, and thus the larger number of ionizing
photons. Therefore, the SFRs for less massive galaxies can be
overestimated due to the metallicity effect, if they follow the
mass-metallicity relation \citep[e.g.,][]{Shimakawa2015a}.
These are areas of research for future papers.

\acknowledgments
We thank the anonymous referee for providing constructive comments.
MH, RS, and TS acknowledge support from the Japan Society for the
Promotion of Science (JSPS) through the JSPS Research Fellowship for
Young Scientists. TK acknowledges the financial support in part by a
Grant-in-Aid for the Scientific Research (Nos.\ 21340045 and 24244015)
by the Japanese Ministry of Education, Culture, Sports and Science. 
This letter is based on data collected at Subaru Telescope, which is
operated by the National Astronomical Observatory of Japan, as well as
observations made with the NASA/ESA Hubble Space Telescope, obtained
at the Space Telescope Science Institute, which is operated by the
Association of Universities for Research in Astronomy, Inc., under
NASA contract NAS 5-26555. These HST observations are associated with
programs GO-13291.

{\it Facilities:} \facility{Subaru, HST}.



\begin{thebibliography}{37}
\expandafter\ifx\csname natexlab\endcsname\relax\def\natexlab#1{#1}\fi

\bibitem[{{Bicker} \& {Fritze-v.~Alvensleben}(2005)}]{Bicker2005}
{Bicker}, J., \& {Fritze-v.~Alvensleben}, U. 2005, \aap, 443, L19

\bibitem[{{Bruzual} \& {Charlot}(2003)}]{BC03}
{Bruzual}, G., \& {Charlot}, S. 2003, \mnras, 344, 1000

\bibitem[{{Bundy} {et~al.}(2006){Bundy}, {Ellis}, {Conselice}, {Taylor},
  {Cooper}, {Willmer}, {Weiner}, {Coil}, {Noeske}, \& {Eisenhardt}}]{Bundy2006}
{Bundy}, K., {et~al.} 2006, \apj, 651, 120

\bibitem[{{Calzetti} {et~al.}(2000){Calzetti}, {Armus}, {Bohlin}, {Kinney},
  {Koornneef}, \& {Storchi-Bergmann}}]{Calzetti2000}
{Calzetti}, D., {Armus}, L., {Bohlin}, R.~C., {Kinney}, A.~L., {Koornneef}, J.,
  \& {Storchi-Bergmann}, T. 2000, ApJ, 533, 682

\bibitem[{{Chabrier}(2003)}]{Chabrier2003}
{Chabrier}, G. 2003, PASP, 115, 763

\bibitem[{{Cooke} {et~al.}(2014){Cooke}, {Hatch}, {Muldrew}, {Rigby}, \&
  {Kurk}}]{Cooke2014}
{Cooke}, E.~A., {Hatch}, N.~A., {Muldrew}, S.~I., {Rigby}, E.~E., \& {Kurk},
  J.~D. 2014, \mnras, 440, 3262

\bibitem[{{Cowie} {et~al.}(1996){Cowie}, {Songaila}, {Hu}, \&
  {Cohen}}]{Cowie1996}
{Cowie}, L.~L., {Songaila}, A., {Hu}, E.~M., \& {Cohen}, J.~G. 1996, \aj, 112,
  839

\bibitem[{{Daddi} {et~al.}(2007){Daddi}, {Dickinson}, {Morrison}, {Chary},
  {Cimatti}, {Elbaz}, {Frayer}, {Renzini}, {Pope}, {Alexander}, {Bauer},
  {Giavalisco}, {Huynh}, {Kurk}, \& {Mignoli}}]{Daddi2007}
{Daddi}, E., {et~al.} 2007, ApJ, 670, 156

\bibitem[{{Dopita} {et~al.}(2006){Dopita}, {Fischera}, {Sutherland}, {Kewley},
  {Tuffs}, {Popescu}, {van Breugel}, {Groves}, \& {Leitherer}}]{Dopita2006}
{Dopita}, M.~A., {et~al.} 2006, \apj, 647, 244

\bibitem[{{Dunne} {et~al.}(2009){Dunne}, {Ivison}, {Maddox}, {Cirasuolo},
  {Mortier}, {Foucaud}, {Ibar}, {Almaini}, {Simpson}, \& {McLure}}]{Dunne2009}
{Dunne}, L., {et~al.} 2009, \mnras, 394, 3

\bibitem[{{Elbaz} {et~al.}(2007){Elbaz}, {Daddi}, {Le Borgne}, {Dickinson},
  {Alexander}, {Chary}, {Starck}, {Brandt}, {Kitzbichler}, {MacDonald},
  {Nonino}, {Popesso}, {Stern}, \& {Vanzella}}]{Elbaz2007}
{Elbaz}, D., {et~al.} 2007, \aap, 468, 33

\bibitem[{{Garn} \& {Best}(2010)}]{GarnBest2010}
{Garn}, T., \& {Best}, P.~N. 2010, \mnras, 409, 421

\bibitem[{{Hayashi} {et~al.}(2012){Hayashi}, {Kodama}, {Tadaki}, {Koyama}, \&
  {Tanaka}}]{Hayashi2012}
{Hayashi}, M., {Kodama}, T., {Tadaki}, K.-I., {Koyama}, Y., \& {Tanaka}, I.
  2012, \apj, 757, 15

\bibitem[{{Ichikawa} {et~al.}(2006){Ichikawa}, {Suzuki}, {Tokoku}, {Uchimoto},
  {Konishi}, {Yoshikawa}, {Yamada}, {Tanaka}, {Omata}, \&
  {Nishimura}}]{ichikawa2006}
{Ichikawa}, T., {et~al.} 2006, in Society of Photo-Optical Instrumentation
  Engineers (SPIE) Conference Series, Vol. 6269, Society of Photo-Optical
  Instrumentation Engineers (SPIE) Conference Series

\bibitem[{{Karim} {et~al.}(2011){Karim}, {Schinnerer},
  {Mart{\'{\i}}nez-Sansigre}, {Sargent}, {van der Wel}, {Rix}, {Ilbert},
  {Smol{\v c}i{\'c}}, {Carilli}, {Pannella}, {Koekemoer}, {Bell}, \&
  {Salvato}}]{Karim2011}
{Karim}, A., {et~al.} 2011, \apj, 730, 61

\bibitem[{{Kennicutt}(1998)}]{kennicutt1998}
{Kennicutt}, Jr., R.~C. 1998, ARA\&A, 36, 189

\bibitem[{{Kodama} {et~al.}(1998){Kodama}, {Arimoto}, {Barger}, \&
  {Arag'on-Salamanca}}]{Kodama1998}
{Kodama}, T., {Arimoto}, N., {Barger}, A.~J., \& {Arag'on-Salamanca}, A. 1998,
  A\&A, 334, 99

\bibitem[{{Kodama} {et~al.}(2013){Kodama}, {Hayashi}, {Koyama}, {Tadaki},
  {Tanaka}, \& {Shimakawa}}]{Kodama2013}
{Kodama}, T., {Hayashi}, M., {Koyama}, Y., {Tadaki}, K.-I., {Tanaka}, I., \&
  {Shimakawa}, R. 2013, in IAU Symposium, Vol. 295, IAU Symposium, ed.
  D.~{Thomas}, A.~{Pasquali}, \& I.~{Ferreras}, 74--77

\bibitem[{{Koekemoer} {et~al.}(2011){Koekemoer}, {Faber}, {Ferguson}, {Grogin},
  {Kocevski}, {Koo}, {Lai}, {Lotz}, {Lucas}, {McGrath}, {Ogaz}, {Rajan},
  {Riess}, {Rodney}, {Strolger}, \& et~al.}]{Koekemoer2011}
{Koekemoer}, A.~M., {et~al.} 2011, \apjs, 197, 36

\bibitem[{{Kova{\v c}} {et~al.}(2014){Kova{\v c}}, {Lilly}, {Knobel},
  {Bschorr}, {Peng}, {Carollo}, {Contini}, {Kneib}, {Le F{\'e}vre}, {Mainieri},
  {Renzini}, {Scodeggio}, {Zamorani}, {Bardelli}, {Bolzonella}, {Bongiorno},
  {Caputi}, {Cucciati}, {de la Torre}, {de Ravel}, {Franzetti}, {Garilli},
  {Iovino}, {Kampczyk}, {Lamareille}, {Le Borgne}, {Le Brun}, {Maier},
  {Mignoli}, {Oesch}, {Pello}, {Montero}, {Presotto}, {Silverman}, {Tanaka},
  {Tasca}, {Tresse}, {Vergani}, {Zucca}, {Aussel}, {Koekemoer}, {Le Floc'h},
  {Moresco}, \& {Pozzetti}}]{Kovac2014}
{Kova{\v c}}, K., {et~al.} 2014, \mnras, 438, 717

\bibitem[{{Koyama} {et~al.}(2013){Koyama}, {Smail}, {Kurk}, {Geach}, {Sobral},
  {Kodama}, {Nakata}, {Swinbank}, {Best}, {Hayashi}, \& {Tadaki}}]{Koyama2013b}
{Koyama}, Y., {et~al.} 2013, \mnras, 434, 423

\bibitem[{{Koyama} {et~al.}(2015){Koyama}, {Kodama}, {Hayashi}, {Shimakawa},
  {Yamamura}, {Egusa}, {Oi}, {Tanaka}, {Tadaki}, {Takita}, \&
  {Makiuti}}]{Koyama2015}
---. 2015, \mnras, 453, 879

\bibitem[{{Madau}(1995)}]{Madau1995}
{Madau}, P. 1995, \apj, 441, 18

\bibitem[{{Muzzin} {et~al.}(2013){Muzzin}, {Marchesini}, {Stefanon}, {Franx},
  {McCracken}, {Milvang-Jensen}, {Dunlop}, {Fynbo}, {Brammer}, {Labb{\'e}}, \&
  {van Dokkum}}]{Muzzin2013}
{Muzzin}, A., {et~al.} 2013, \apj, 777, 18

\bibitem[{{Noeske} {et~al.}(2007){Noeske}, {Weiner}, {Faber}, {Papovich},
  {Koo}, {Somerville}, {Bundy}, {Conselice}, {Newman}, {Schiminovich}, {Le
  Floc'h}, {Coil}, {Rieke}, {Lotz}, {Primack}, {Barmby}, {Cooper}, {Davis},
  {Ellis}, {Fazio}, {Guhathakurta}, {Huang}, {Kassin}, {Martin}, {Phillips},
  {Rich}, {Small}, {Willmer}, \& {Wilson}}]{Noeske2007}
{Noeske}, K.~G., {et~al.} 2007, \apjl, 660, L43

\bibitem[{{Oke} \& {Gunn}(1983)}]{Oke1983}
{Oke}, J.~B., \& {Gunn}, J.~E. 1983, \apj, 266, 713

\bibitem[{{Peng} {et~al.}(2012){Peng}, {Lilly}, {Renzini}, \&
  {Carollo}}]{Peng2012}
{Peng}, Y.-j., {Lilly}, S.~J., {Renzini}, A., \& {Carollo}, M. 2012, \apj, 757,
  4

\bibitem[{{Reddy} {et~al.}(2012){Reddy}, {Pettini}, {Steidel}, {Shapley},
  {Erb}, \& {Law}}]{Reddy2012}
{Reddy}, N.~A., {Pettini}, M., {Steidel}, C.~C., {Shapley}, A.~E., {Erb},
  D.~K., \& {Law}, D.~R. 2012, \apj, 754, 25

\bibitem[{{Rodighiero} {et~al.}(2011){Rodighiero}, {Daddi}, {Baronchelli},
  {Cimatti}, {Renzini}, {Aussel}, {Popesso}, {Lutz}, {Andreani}, {Berta},
  {Cava}, {Elbaz}, {Feltre}, {Fontana}, {F{\"o}rster Schreiber},
  {Franceschini}, {Genzel}, {Grazian}, {Gruppioni}, {Ilbert}, {Le Floch},
  {Magdis}, {Magliocchetti}, {Magnelli}, {Maiolino}, {McCracken}, {Nordon},
  {Poglitsch}, {Santini}, {Pozzi}, {Riguccini}, {Tacconi}, {Wuyts}, \&
  {Zamorani}}]{Rodighiero2011}
{Rodighiero}, G., {et~al.} 2011, \apjl, 739, L40

\bibitem[{{Shimakawa} {et~al.}(2015{\natexlab{a}}){Shimakawa}, {Kodama},
  {Tadaki}, {Hayashi}, {Koyama}, \& {Tanaka}}]{Shimakawa2015a}
{Shimakawa}, R., {Kodama}, T., {Tadaki}, K.-i., {Hayashi}, M., {Koyama}, Y., \&
  {Tanaka}, I. 2015{\natexlab{a}}, \mnras, 448, 666

\bibitem[{{Shimakawa} {et~al.}(2014){Shimakawa}, {Kodama}, {Tadaki}, {Tanaka},
  {Hayashi}, \& {Koyama}}]{Shimakawa2014}
{Shimakawa}, R., {Kodama}, T., {Tadaki}, K.-i., {Tanaka}, I., {Hayashi}, M., \&
  {Koyama}, Y. 2014, \mnras, 441, L1

\bibitem[{{Shimakawa} {et~al.}(2015{\natexlab{b}}){Shimakawa}, {Kodama},
  {Steidel}, {Tadaki}, {Tanaka}, {Strom}, {Hayashi}, {Koyama}, {Suzuki}, \&
  {Yamamoto}}]{Shimakawa2015b}
{Shimakawa}, R., {et~al.} 2015{\natexlab{b}}, \mnras, 451, 1284

\bibitem[{{Shivaei} {et~al.}(2015){Shivaei}, {Reddy}, {Shapley}, {Kriek},
  {Siana}, {Mobasher}, {Coil}, {Freeman}, {Sanders}, {Price}, {de Groot}, \&
  {Azadi}}]{Shivaei2015}
{Shivaei}, I., {et~al.} 2015, \apj, 815, 98

\bibitem[{{Sobral} {et~al.}(2013){Sobral}, {Smail}, {Best}, {Geach}, {Matsuda},
  {Stott}, {Cirasuolo}, \& {Kurk}}]{Sobral2013}
{Sobral}, D., {Smail}, I., {Best}, P.~N., {Geach}, J.~E., {Matsuda}, Y.,
  {Stott}, J.~P., {Cirasuolo}, M., \& {Kurk}, J. 2013, \mnras, 428, 1128

\bibitem[{{Speagle} {et~al.}(2014){Speagle}, {Steinhardt}, {Capak}, \&
  {Silverman}}]{Speagle2014}
{Speagle}, J.~S., {Steinhardt}, C.~L., {Capak}, P.~L., \& {Silverman}, J.~D.
  2014, \apjs, 214, 15

\bibitem[{{Suzuki} {et~al.}(2008){Suzuki}, {Tokoku}, {Ichikawa}, {Uchimoto},
  {Konishi}, {Yoshikawa}, {Tanaka}, {Yamada}, {Omata}, \&
  {Nishimura}}]{suzuki2008}
{Suzuki}, R., {et~al.} 2008, PASJ, 60, 1347

\bibitem[{{Whitaker} {et~al.}(2014){Whitaker}, {Franx}, {Leja}, {van Dokkum},
  {Henry}, {Skelton}, {Fumagalli}, {Momcheva}, {Brammer}, {Labb{\'e}},
  {Nelson}, \& {Rigby}}]{Whitaker2014}
{Whitaker}, K.~E., {et~al.} 2014, \apj, 795, 104

\end{thebibliography}
\end{document}